\documentclass[lettersize,journal]{IEEEtran}
\usepackage{amsmath,amsfonts}
\usepackage{algorithmic}
\usepackage{algorithm}
\usepackage{array}
\usepackage[caption=false,font=normalsize,labelfont=sf,textfont=sf]{subfig}
\usepackage{textcomp}
\usepackage{stfloats}
\usepackage{url}
\usepackage{verbatim}
\usepackage{graphicx}
\usepackage{cite}
\begin{document}

\title{Wi-Fi 8: Embracing the Millimeter-Wave Era}

\author{Xiaoqian Liu, Tingwei Chen, Yuhan Dong, Zhi Mao, Ming Gan, Xun Yang and Jianmin Lu

}



\maketitle

\begin{abstract}
With the increasing demands in communication, Wi-Fi technology is advancing towards its next generation. As high-need applications like Virtual Reality (VR) and Augmented Reality (AR) emerge, the role of millimeter-wave (mmWave) technology becomes critical. This paper explores Wi-Fi 8's potential features, especially its integration of mmWave technology. We address the challenges of implementing mmWave under current protocols and examine the compatibility of new features with mmWave. Our study includes system-level simulations, upclocking the 802.11ac PPDU to 60 GHz, and considers hardware limitations. The results demonstrate significant performance improvements with mmWave in Wi-Fi 8, indicating its feasibility for high-demand wireless scenarios.
\end{abstract}

\begin{IEEEkeywords}
Wi-Fi 8, mmWave, feature, hardware impairment.
\end{IEEEkeywords}

\section{Introduction}

Wi-Fi technology has evolved significantly to meet the demands of an increasingly connected world, supporting applications that require low latency, high throughput, and seamless connectivity. From Wi-Fi 1 to Wi-Fi 7, operating bands have expanded from 2.4 GHz to 6 GHz. However, as \cite{website6GHz} notes, differing global regulations on the use of the 6 GHz band, with some regions allocating this band to 5G, have limited the potential advantages of Wi-Fi 7's large bandwidth. Given this situation, the exploration of millimeter wave (mmWave) frequency bands emerges as a significant  opportunity to recapture the benefits initially earmarked for the 6 GHz band.

By 2028, we anticipate an accelerated need for fresh spectrums due to the exponential growth of Internet of Things devices and the rise of applications demanding high throughput and low latency. Further exploration of wider bandwidths in the 6 GHz range may not provide substantial additional value due to regional bandwidth restrictions.
While it is plausible to adjust and improve the protocol to integrate more features to meet our needs, the unlocking of frequency bands beyond 6 GHz could be a more fruitful approach. Under these circumstances, the 45 GHz and 60 GHz mmWave unlicensed band holds considerable promise, offering a rich 14 GHz spectrum. This band brings high transmission rates, substantial information capacity, and easy implementation of narrow beamwidth and high gain antennas, leading to high resolution and excellent interference resistance. The Wireless Gigabit (WiGig) standard, known as 60 GHz Wi-Fi and the emergence of 5G's Frequency Range 2 (FR2),  the operating band spanning from 26.5 GHz to 71.00 GHz, indicate that the advancement of mmWave hardware technology is in progress. As these technologies become mainstream and start seeing commercial usage, the cost concerns surrounding mmWave communication will likely diminish with falling hardware costs.
Looking towards the future, we can foresee multi-link devices (MLDs) allowing the coexistence of mmWave bands with lower frequency bands\cite{Cariou2022}. This evolution builds on the current multi-link framework. While single access point (AP) scenarios present relatively simple deployment opportunities, designing the medium access control (MAC) and physical (PHY) layers of Wi-Fi products for Multi-AP coordination (MAPC) has yet to be fully explored.

Besides, a significant challenge lies in verifying the feasibility of mmWave technology under the current sub-7 GHz protocols. In fact, there are several IEEE 802.11 standards, such as 802.11ad \cite{nitsche2014ieee} and 802.11ay \cite{ghasempour2017ieee}, which utilize mmWave technology. To better understand the innovative strides made in Wi-Fi 8, we will examine the distinctive differences and enhancements in mmWave technology as compared to the existing standards like 802.11ad and 802.11ay:
\begin{itemize}
    \item \textbf{Limitations in 802.11ad and 802.11ay's mmWave:} While 802.11ad and 802.11ay were pioneers in mmWave technology, their designs were exclusively tailored for mmWave communication. This focus resulted in a notable gap in integrating high-frequency mmWave bands with the lower frequency spectrum and consequently led to increased development and hardware costs. The need for distinct physical layer devices and algorithms specific to mmWave frequencies further emphasized this design limitation.
    
    \item \textbf{Wi-Fi 8's reuse of lower frequency devices and algorithms:} Wi-Fi 8 innovates by fully reusing lower frequency physical layer devices and algorithms, optimizing the development process, and reducing costs.
    
    \item \textbf{Bandwidth strategy in Wi-Fi 8:} Wi-Fi 8 operates at a minimum of 160 MHz or 320 MHz, a decrease from the typical 2.16 GHz bandwidth in earlier mmWave standards. This is more cost-effective and suitable for enterprise scenarios.
    
    \item \textbf{Improved Network Efficiency with Wi-Fi 8:} The shift to lower operational bandwidths in Wi-Fi 8 allows for improved frequency division multiplexing, moving from 20/40 MHz to 320 MHz, better supporting densely networked environments.
\end{itemize}

In the era of Wi-Fi 8, a natural progression is the incorporation of mmWave bands, a shift that presents unique challenges, particularly in integrating mmWave technology with existing sub-7 GHz protocols. This integration can lead to specific hardware impairments, thus requiring thorough experimental validation. Moreover, the potential transition of certain deferred features from Wi-Fi 7 to Wi-Fi 8 adds another layer of complexity, emphasizing the need to assess their compatibility with mmWave technology. Building upon prior studies of new Wi-Fi 8 features \cite{reshef2022future, giordano2023will}, this paper delves into exploring these potential features, closely following the recommendations of the IEEE 802.11 Integrated Millimeter Wave Study Group (IMMW SG) with a focus on mmWave technology. We commence by examining candidate features for Wi-Fi 8, including those carried over from Wi-Fi 7, and then shift to discussing the integration of mmWave technology within the Multi-Link Operation (MLO) framework. The final section of our study involves upclocking the carrier frequency of the 802.11ac physical protocol data unit (PPDU) to 60 GHz, taking into consideration the implications of traditional hardware limitations through a system-level simulation. Our findings ultimately demonstrate the feasibility of mmWave communication technology in Wi-Fi operational scenarios, providing insights into the future landscape of wireless networking.


\section{Wi-Fi Generations}
Since IEEE 802.11-1997 standardized wireless local area network (WLAN), Wi-Fi has begun to develop.
Later, after revision of IEEE 802.11a, 802.11b and 802.11g, Wi-Fi technology has gradually been improved in terms of throughput and frequency bands to meet various requirements of industry and scientific research \cite{reshef2022future}.

\begin{figure*}[!t]
	\centering
	\includegraphics[width=7.30in]{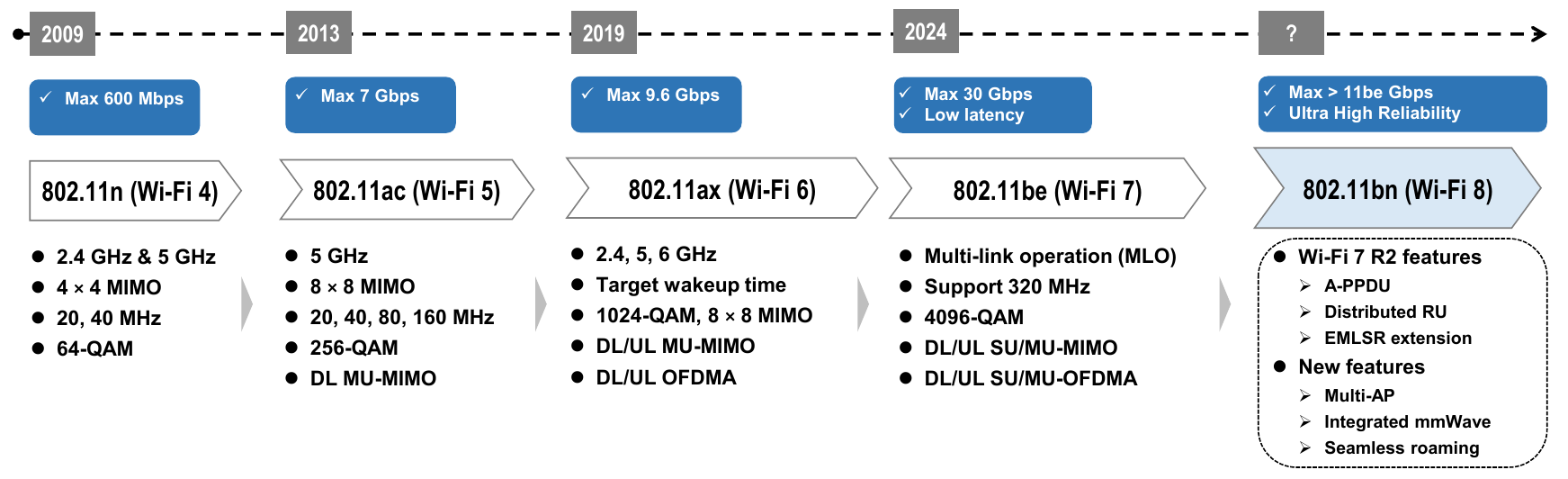}
	\caption{Illustration of Wi-Fi technology evolution.}
	\label{fig_1}
\end{figure*}

Fig.~\ref{fig_1} shows the evolution from Wi-Fi 4 to Wi-Fi 8. 
Wi-Fi 4 was released in 2009 and based on IEEE 802.11n standard. It supports both 2.4 GHz and 5 GHz radio bands and can provide max data rate up to 600 Mbps.
In addition, Wi-Fi 4 introduces the multiple-input multiple-output (MIMO) communication technology for the first time.
Wi-Fi 5 is IEEE 802.11ac standard, released in late 2013.
To promote the development of high frequency band Wi-Fi technologies, Wi-Fi 5 supports only 5 GHz. It supports larger channel bandwidth and 256-quadrature amplitude modulation (QAM), so it more than doubles the theoretical maximum data rate of Wi-Fi 4.
In 2019, Wi-Fi 6 based on IEEE 802.11ax entered the market and introduced a major paradigm shift in Wi-Fi technology.
In all prior Wi-Fi generations to Wi-Fi 6, each device needs to independently contend for channel access, but Wi-Fi 6 allows an AP to trigger multiple stations (STAs) to transmit data at the same time.
It is a trigger-based access and enables uplink multi-user (MU) operation.
Wi-Fi 6 supports both uplink (UL) and downlink (DL) MU-MIMO, whereas Wi-Fi 5 supports only DL MU-MIMO. In addition, Wi-Fi 6 introduces orthogonal frequency-division multiple access (OFDMA) and uses this technology to implement DL and UL multi-user operations.
The technology innovation of Wi-Fi 6 also includes 1024-QAM, target wakeup time (TWT) and so on.

The final amendments of the above standards have been released, however, Wi-Fi 7 and Wi-Fi 8 are still under study.
Most of the work on Wi-Fi 7 has been done and the final amendment is scheduled to be released in December 2024.
Wi-Fi 7 corresponds to IEEE 802.11be - Extremely High Throughput (EHT) and introduces 320 MHz bandwidth, 4096-QAM, multiple resource unit (MRU), MLO, low-latency operations and so on \cite{garcia2021ieee,deng2020ieee}.
Wi-Fi 8 is just beginning to be explored, and it is based on IEEE 802.11bn Ultra High Reliability (UHR).
Wi-Fi 8 will inherit some leftover features from Wi-Fi 7, such as aggregated PPDU (A-PPDU) and distributed RU, and expand some new functions, the most important of which is integrated mmWave. 
Expected new features brought about by Wi-Fi 8 are covered in the next section.

\section{Wi-Fi 8 potential features}
Considering the high complexity that some candidate features could introduce, several initially considered for Wi-Fi 7 have not been incorporated. Based on past experiences, they might be deferred to the next generation. Hence, these leftover features from Wi-Fi 7 are likely to be included in Wi-Fi 8. We should thus prioritize them in our discussion. In the following sections, we will focus our discussion on these potential Wi-Fi 8 candidate features, specifically those that are compatible with and can leverage mmWave technology.

This includes addressing the challenges in high-density Wi-Fi scenarios, where numerous STAs connect to APs in confined spaces. These scenarios introduce challenges when using mmWave, known for limited coverage. To address these challenges in Wi-Fi 8, features such as MAPC and seamless roaming will be essential for maintaining reliable and high-quality connectivity.

\subsection{Multi-AP coordination}
When multiple APs are deployed in a network environment to extend the network's range and capacity, they can also introduce some interference. MAPC is a key feature of Wi-Fi 8, aiming to optimize channel selection and balance loads between APs to ensure effective use and fair sharing of radio resources. This approach allows multiple APs to collaborate and share spectrum resources, reducing unnecessary contention and improving network efficiency. The combination of MAPC and mmWave in Wi-Fi 8 offers a vast array of opportunities and possibilities\cite{verma2023survey}. However, existing protocols for APs coordination are not without limitations, often leading to suboptimal performance. 
To address these challenges, a design of the coordination protocols are necessary. To achieve this, MAPC will incorporate various techniques, including joint transmission (JT), coordinated beamforming (Co-BF), coordinated UL MU MIMO (Co-UL-MU-MIMO), coordinated spatial reuse (Co-SR), coordinated OFDMA (Co-OFDMA), and coordinated time division multiple access (Co-TDMA). 

While MAPC in Wi-Fi 8 brings together a suite of advanced techniques to address these coordination challenges, one particularly effective solution is Co-SR. According to the simulation in \cite{nunez2023multi}, Co-SR allows MAPs to transmit simultaneously in the same slot, improving throughput and reducing delay compared to Co-TDMA. Being an extension of the current spatial reuse scheme in TGax, Co-SR avoids the need for a multi-AP sounding procedure, achieving a balance between performance gain and implementation complexity. Given its low complexity, Co-SR stands as a promising candidate feature for MAPC.

\subsection{Seamless roaming}
Firstly, it is important to note the emergence of the new MLO architecture in Wi-Fi 7. To differentiate it from previous architectures, we redefine the classic AP and STA as AP MLD and STA MLD, respectively. These terms refer to APs and STAs equipped to support MLD, with AP MLDs sharing a common MAC layer.

\begin{figure*}[!t]
	\centering
	\includegraphics[width=7.30in]{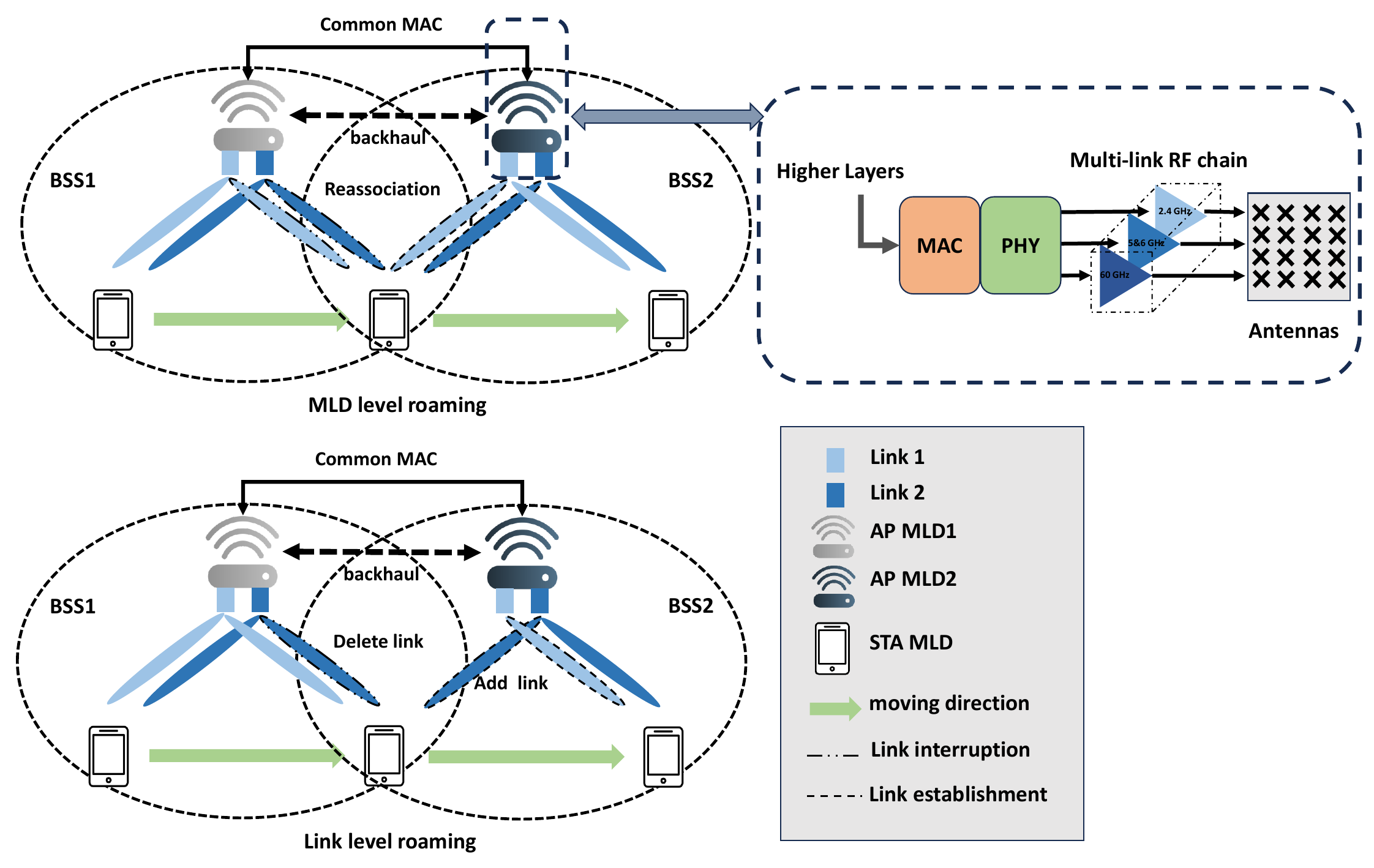}
	\caption{Comparative Overview of MLD Level Roaming, Link Level Roaming, and Multi-Link Architecture}
	\label{fig_8}
\end{figure*}

As illustrated in Fig.~\ref{fig_8}, the roaming technology in Wi-Fi 7, known as MLD-level roaming, requires that the STA MLD must disassociate from the current AP MLD  and reassociate with a new AP MLD during the roaming process. However, this process requires interrupting the transmission of data frames, which is unacceptable in UHR scenarios. Therefore, maintaining the STA connection while roaming between different BSSs is crucial for ensuring the quality of service (QoS). A potential solution is to implement link-level roaming. When roaming occurs, the STA first deletes a higher-frequency link, maintaining a connection via a lower-frequency link to continue data transmission. Subsequently, it adds a link to the approaching AP MLD2, then disconnects all links from AP MLD1 and fully connects to AP MLD2, achieving seamless roaming. In this approach,  non-AP STA MLDs can simultaneously utilize links from multiple AP MLDs for frame exchange \cite{Liwen2023}. This ensures that there is at least one active link during the STA transition, enhancing the reliability of roaming.

With the incorporation of mmWave frequencies, the options for seamless roaming are expanded, further stabilizing communication during the roaming process.

\begin{table*}[t] 
	\renewcommand\arraystretch{1.4}
	\begin{center}
		\caption{An upclocked PPDU version}
		\label{table_3}
		\setlength{\tabcolsep}{3 mm}{
			\begin{tabular}{|c|c|c|c|}
				\hline   \rule{0pt}{8pt} ~ & \textbf{4x 802.11ac} & \textbf{8x 802.11ac} & \textbf{16x 802.11ac}  \\   
				\hline   \rule{0pt}{8pt} \textbf{Bandwidth} & 80/160/320/640 MHz & 160/320/640/1280 MHz & 320/640/1280/2560/5120 MHz  \\ 
				\hline   \rule{0pt}{8pt} \textbf{Subcarrier spacing} & 1.25 MHz & 2.5 MHz & 1.25 MHz (802.11be part)   \\ 
				\hline   \rule{0pt}{8pt} \textbf{IDFT/DFT period} & 800 ns & 400 ns & 800 ns   \\   
				\hline   \rule{0pt}{8pt} \textbf{IDFT/DFT sive} & 64/128/256/512 & 64/128/256/512 & 256/512/1024/2048/4096  \\ 
				\hline
		\end{tabular} }
	\end{center}
\end{table*}

\section{Integrated Millimeter Wave Technology}
The study group dedicated to integrated mmWave technology is set to commence its work in November 2023\cite{deVegt2023}. The group's research will be focused on exploring the potential applications of mmWave within the forthcoming Wi-Fi 8 standard. The significance of this technology could lead to a unique certification, potentially known as Wi-Fi 8E, emphasizing the crucial role of mmWave technology in the future of wireless communication. Wi-Fi x (e.g., Wi-Fi 7) is named by Wi-Fi alliance. Although IMMW is going to be independent task group from UHR (802.11bn) in IEEE 802.11, IMMW is still expected to be certified as Wi-Fi  8E while UHR (802.11bn) will be certified as the first version of Wi-Fi 8 in Wi-Fi alliance.


The mmWave technology operates at high frequencies (30-300 GHz) and can support high data rates, making it suitable for various wireless communication applications. 
The most important benefit of mmWave is high throughput and low latency.
In addition, high frequency communication will naturally reduce the delay of each communication and allow for a substantial increase in frequency division multiplexing.
In this way, each link can individually use a portion of the spectrum without interfering with each other and QoS is guaranteed in very dense environments.
mmWave will also address the lack of 6 GHz band for Wi-Fi in some countries, providing a strong lighthouse client feature and solutions for the development of integrated communications and sensing.

It should be noted that the usage scope of mmWave needs to be clearly defined in the UHR.
In fact, as mentioned above, simply operating in mmWave band can bring a lot of benefits and low complexity and cost such that it could be commercialized soon, so there is no need to define more additional functions, and related optimizations can be defined in future Wi-Fi generations.
For example, we believe that only single-user transmissions are required without defining multi-user operations.
In mmWave operations, the components that need to be design include simple beamforming training sequence, minor adaptations to MLO, and PHY designs for different bandwidth OFDM operations which are implemented in sub-7 GHz Wi-Fi.
A good direction for PHY design is to reuse the lower band as much as possible by using upclocking.
The existing low band Wi-Fi solutions can be upclocked to the mmWave band to minimize complexity and facilitate rapid implementation.
Besides, reusing legacy algorithms and architectures can help save algorithm design and validation efforts.
A simple example is that a mmWave PHY can be implemented by increasing the subcarrier spacing and channel bandwidth of IEEE 802.11ac by a specific multiple.
Correspondingly, the OFDM symbol time is also reduced, but the baseband processing remains unchanged.
A feasible mmWave PHY design is shown in Table~\ref{table_3}.

In order to integrate 60 GHz, the architecture of multi-link can be used, which is shown in Fig.~\ref{fig_8}.
The use of MLO has a number of benefits \cite{chen2022ieee}.
With MLO, discovery, association, scheduling, and broadcast exchanging can be done in lower band.
During beamforming training, sector sweep is performed in the 60 GHz band, but the sequences can be triggered or scheduled from a lower band, and feedback can be provided in lower band as well.
In addition, devices can seamlessly and quickly fall back to the lower band in case of 60 GHz link breaks.
Although beamforming training needs to be performed repeatedly, the device can resist blockage in the mmWave band.
Power save is also an important issue for mmWave.
Using the target wake time service period (TWT SP) on mmWave band will significantly help power consumption since there is hardly any contention required except to wake up for TWT SP. Through the MLO, TWT scheduling can be negotiated in lower band.
In other words, all broadcasted exchanges, as well as all exchanges done prior to an established and successful beamforming training in the mmWave band, can be done in the lower band \cite{MLO2022}.
Therefore, there is no need to define the mode for control PHY, which greatly simplifies the standard design and implementation.
In summary, we believe that all devices operating in mmWave should have MLO capability.
In addition to the STA/AP in the mmWave band, there should be at least one affiliated STA/AP in the lower band.

Compared to 802.11ad and 802.11ay, specific differences in the next generation of Wi-Fi standards regarding mmWave support include the full reuse of lower band modems, no need for a control PHY, and the use of the MLO framework


There is ongoing research and development focused on probing the further applications of integrated mmWave technology within the IEEE 802.11 standards. This continued exploration of mmWave's potential presents a promising prospect and is anticipated to become a key feature of the forthcoming Wi-Fi 8 standard.

\begin{table}[t]
	\renewcommand\arraystretch{1.4}
	\begin{center}
		\caption{Simulation parameters for performance verification}
		\label{table_2}
		\setlength{\tabcolsep}{3 mm}{
			\begin{tabular}{|c|c|}
				\hline   \rule{0pt}{8pt} \textbf{Parameter} & \textbf{Description}  \\   
				\hline   \rule{0pt}{8pt} MCS & MCS 0, MCS 1 and MCS 4  \\ 
				\hline   \rule{0pt}{8pt} PPDU type & VHT format   \\ 
				\hline   \rule{0pt}{8pt} Channel model & 802.11ad channel model   \\   
				\hline   \rule{0pt}{8pt} TX/RX & 1T1R  \\ 
				\hline   \rule{0pt}{8pt} $f_{c}$ & 73.44 GHz  \\ 
				\hline   \rule{0pt}{8pt} Packet size & 4096 bytes  \\ 
				\hline   \rule{0pt}{8pt} Other considerations & CFO, PN and PA  \\ 
				\hline
		\end{tabular} }
	\end{center}
\end{table}

\section{Performance Evaluation of mmWave}
mmWave is an effective solution to improve throughput and latency performance, but high frequency bands communications are easily affected by device characteristics.
In this section, we briefly introduce the conventional hardware constraints, including carrier frequency offset (CFO), phase noise (PN), and nonlinearities in power amplifiers (PA) and discuss their effects on high frequency communications through simulations.

\subsection{Simulation configuration}
The IEEE 802.11 standards use orthogonal frequency division multiplexing (OFDM) as the modulation scheme, which has the characteristics of anti-multipath fading. 
However, compared to single-carrier systems, the existence of CFO will introduce severe inter-carrier interference (ICI) for OFDM, which needs to be compensated by some CFO estimation algorithms \cite{CFO}.
The generation of PN is related to the performance of the oscillator itself and can be interpreted as a parasitic phase modulation.
PN needs to be carefully considered in 60 GHz OFDM systems because PN is more severe at high frequencies and affects system performance significantly.
Therefore, it is important to predict the tolerable phase noise and use reliable phase correction mechanisms.
The OFDM signal is distorted after being processed by nonlinear high-power amplifier (HPA), which affects the bit error ratio (BER) performance of the system. 
The main reason for this phenomenon is that in order to ensure high power amplification efficiency, the HPA must work near the saturation point as much as possible. 


In order to verify the impact of above hardware impairments on 60 GHz communication performance, we designed system-level simulation.
The simulation discussed the upclocked PPDU based on IEEE 802.11ac and considered the effects of CFO, PN and PA.
In order to better illustrate the impact of hardware impairments, we refrained from using estimation and compensation algorithms for CFO and PN. Additionally, this simulation employs a single-antenna system without utilizing beamformed transmission.
The main purpose of simulation is to verify whether an upclocked sub-7 GHz PPDU version (such as IEEE 802.11ac) works well in a mmWave scenario and the simulation parameters are shown in Table~\ref{table_2}. 
The modulation and coding scheme (MCS) is set to 0, 1, and 4, indicating binary phase shift keying (BPSK), quadrature phase shift keying (QPSK), and 16-QAM respectively.
The PPDU type is very high throughput (VHT) format, which is used in IEEE 802.11ac. The packet size is 4096 bytes.
We consider the conference room (CR) channel model in IEEE 802.11ad standard and the number of transmitters (TX) and receivers (RX) is 1.
$f_{c}$ is center frequency of the carrier, which is the channel 8 defined in IEEE 802.11ay.


A larger subcarrier spacing will reduce the impact of CFO. If an auto-correlation detector is used, the short training field (STF) and long training field (LTF) sequences in sub-7 GHz can resolve a frequency offset up to ± 625 KHz and ± 156.25 KHz, respectively.
In our simulation, the oscillator inaccuracy on both TX and RX sides are 20 part per million (ppm), so the system should compensate for a total of 40 ppm CFO.
For a 73.44 GHz system we set up, the CFO is up to ± 2.94 MHz. Therefore, the subcarrier spacing of at least one training field should be larger than 5.88 MHz.
Ignoring the flicker noise and the reference oscillator noise, we use 1 pole, 1 zero phase noise model. The pole frequency $f_{p}$ and the zero frequency $f_{z}$ is set to 1 MHz and 100 MHz, respectively and the zero point of power spectral density (PSD) is -93 dBc/Hz.
The saturation region of the power amplifier will affect the output signal, and the output backoff (OBO) can be used to mitigate the impact. 
In this work, PA OBO is set to 8 dB and the PA non-linearity model is modified Rapp model, which can describe HPA nonlinear relationship between output power and input power (AM-AM effect) and output phase and input power (AM-PM effect).

\subsection{Simulation results}
Fig. shows the packet error ratio (PER) performance considering hardware impairments (CFO, PN and PA) and without considering these non-idealities.
It can be seen obviously that when 4$\times$ upclocking is used, the PER is 1 regardless of the MCS selection, indicating that data transmission fails.
The conventional subcarrier spacing of Non-HT STF (L-STF) is 4 $\times$ 312.5 KHz, so the threshold for CFO correction can be obtained by dividing the minimum subcarrier spacing 5.88 MHz by the 1250 KHz.
Therefore, only a larger upclocking version than 4.7 can rectify the CFO correctly, which is also reflected in Fig., because both 8$\times$ upclocking and 16$\times$ upclocking achieve better performance under each MCS.
It can be observed that when not considering non-idealities, the system performance is better across various configurations, especially in low-order MCS scenarios.
The smaller the MCS, the lower the SNR required to achieve the specified PER ($10^{-2}$), and the system performance is better. It is intuitive because the low-order MCS brings larger constellation distance and better decoding performance.
In addition, the PER performance improves with the increase of subcarrier spacing for high MCSs, but for low MCSs, the PER performance does not change significantly.
Fig. illustrates the constellation diagram at the RX for the scenario with SNR of 20 dB and MCS of 4. It can be observed that non-idealities have a noticeable impact on the received signal, ultimately leading to an increase in PER.

\subsection{Discussion}

Through the above analysis, we can see that although the hardware impairments will affect the communication system theoretically, the performance deterioration caused by these impairments is still acceptable from the simulation results.

By using larger subcarrier spacing, the impact of CFO can be minimized. 
Therefore, a suitable upclocked version of IEEE 802.11ac PPDU can be used to implement mmWave communications.
In addition, the CFO can be estimated and compensated, but this is an algorithmic issue and we don't discuss it much in the standard.
PN usually causes ICI and common phase error (CPE), but these effects can be overcome by proper pilot designs.
PA OBO is an important method to overcome PA nonlinearities, and the selection of PA backoff needs to be carefully considered.
The IEEE 802.11ay standard has specified a value of PA backoff in high frequency communication.
Similarly, in the future Wi-Fi 8 standard, selecting an appropriate PA backoff value will be an important issue.
In fact, not only CFO, PN, and PA nonlinearities, but also other hardware impairments (such as I/Q imbalance and so on) that affect communications, which need to be discussed in the next generation of standard.

mmWave communication is not difficult to achieve.
In fact, the main issue seems to be the balance between the additional cost and gain.
mmWave is a promising feature helping reach the UHR objectives. 
We believe there are enough gains to implement mmWave today, and it’s the time to have it now.

\section{Conclusion}
Wi-Fi has always been one of the most important wireless technologies today, providing a great impetus for the development of socio-economic and scientific research.
Since Wi-Fi was defined, researchers have been continuously innovating Wi-Fi to improve performance in terms of data rate, access efficiency, and latency to meet increasing user requirements.

In this paper, we first introduce the technical features of each generation of Wi-Fi and describe the possible innovations of Wi-Fi 8.
We then covered the features of Wi-Fi 8 in detail, including leftover features from Wi-Fi 7 and some new promising features.
In our view, integrated mmWave will be a key technology of Wi-Fi 8. 
We present a system-level simulation to demonstrate that mmWave can overcome the effects of high-frequency hardware impairment.
Wi-Fi 8 is an expected technology. In the future, research on Wi-Fi 8 will be gradually promoted to meet the requirements of high reliability, low latency, and high throughput.




%

\bibliographystyle{IEEEtran}
\bibliography{IEEEabrv,reference}

\begin{thebibliography}{10}
\providecommand{\url}[1]{#1}
\csname url@samestyle\endcsname
\providecommand{\newblock}{\relax}
\providecommand{\bibinfo}[2]{#2}
\providecommand{\BIBentrySTDinterwordspacing}{\spaceskip=0pt\relax}
\providecommand{\BIBentryALTinterwordstretchfactor}{4}
\providecommand{\BIBentryALTinterwordspacing}{\spaceskip=\fontdimen2\font plus
\BIBentryALTinterwordstretchfactor\fontdimen3\font minus \fontdimen4\font\relax}
\providecommand{\BIBforeignlanguage}[2]{{%
\expandafter\ifx\csname l@#1\endcsname\relax
\typeout{** WARNING: IEEEtran.bst: No hyphenation pattern has been}%
\typeout{** loaded for the language `#1'. Using the pattern for}%
\typeout{** the default language instead.}%
\else
\language=\csname l@#1\endcsname
\fi
#2}}
\providecommand{\BIBdecl}{\relax}
\BIBdecl

\bibitem{website6GHz}
{Wi-Fi Alliance}, ``Countries enabling wi-fi in 6 ghz (wi-fi 6e),'' July 2023, https://www.wi-fi.org/countries-enabling-wi-fi-in-6-ghz-wi-fi-6e, accessed Aug. 14, 2023.

\bibitem{Cariou2022}
L.~Cariou, R.~Stacey, and C.~Cordeiro, ``{mmWave operation in UHR},'' Nov 2022, {IEEE} 802.11-22/1884r0.

\bibitem{nitsche2014ieee}
T.~Nitsche, C.~Cordeiro, A.~B. Flores, E.~W. Knightly, E.~Perahia, and J.~C. Widmer, ``{IEEE 802.11 ad: directional 60 GHz communication for multi-Gigabit-per-second Wi-Fi},'' \emph{IEEE Commun. Mag.}, vol.~52, no.~12, pp. 132--141, 2014.

\bibitem{ghasempour2017ieee}
Y.~Ghasempour, C.~R. Da~Silva, C.~Cordeiro, and E.~W. Knightly, ``{IEEE 802.11 ay: Next-generation 60 GHz communication for 100 Gb/s Wi-Fi},'' \emph{IEEE Commun. Mag.}, vol.~55, no.~12, pp. 186--192, 2017.

\bibitem{reshef2022future}
E.~Reshef and C.~Cordeiro, ``Future directions for wi-fi 8 and beyond,'' \emph{IEEE Commun. Mag.}, vol.~60, no.~10, pp. 50--55, 2022.

\bibitem{giordano2023will}
L.~G. Giordano, G.~Geraci, M.~Carrascosa, and B.~Bellalta, ``What will wi-fi 8 be? a primer on ieee 802.11 bn ultra high reliability,'' \emph{arXiv preprint arXiv:2303.10442}, 2023.

\bibitem{garcia2021ieee}
A.~Garcia-Rodriguez, D.~Lopez-Perez, L.~Galati-Giordano, and G.~Geraci, ``{IEEE 802.11 be: Wi-Fi 7 strikes back},'' \emph{IEEE Commun. Mag.}, vol.~59, no.~4, pp. 102--108, 2021.

\bibitem{deng2020ieee}
C.~Deng, X.~Fang, X.~Han, X.~Wang, L.~Yan, R.~He, Y.~Long, and Y.~Guo, ``{IEEE 802.11 be Wi-Fi 7: New challenges and opportunities},'' \emph{IEEE Communications Surveys \& Tutorials}, vol.~22, no.~4, pp. 2136--2166, 2020.

\bibitem{verma2023survey}
S.~Verma, T.~K. Rodrigues, Y.~Kawamoto, and N.~Kato, ``{A Survey on Multi-AP Coordination Approaches over Emerging WLANs: Future Directions and Open Challenges},'' \emph{arXiv preprint arXiv:2306.04164}, 2023.

\bibitem{nunez2023multi}
D.~Nunez, M.~Smith, and B.~Bellalta, ``Multi-ap coordinated spatial reuse for wi-fi 8: Group creation and scheduling,'' \emph{arXiv preprint arXiv:2305.04846}, 2023.

\bibitem{Liwen2023}
L.~Chu, K.~Ryu, H.~Wang, h.~Zhang, R.~Cao, Y.~Zhang, S.~Srinivasa, and H.-L. Lou, ``{Smooth Roaming},'' January 2023, {IEEE} 802.11-23/0170r1.

\bibitem{deVegt2023}
R.~de~Vegt, ``{Integrated mmWave Study Group Creation},'' March 2023, {IEEE} 802.11-23/0481r2.

\bibitem{chen2022ieee}
C.~Chen, X.~Chen, D.~Das, D.~Akhmetov, and C.~Cordeiro, ``{Overview and Performance Evaluation of Wi-Fi 7},'' \emph{IEEE Commun. Standards Mag.}, vol.~6, no.~2, pp. 12--18, 2022.

\bibitem{MLO2022}
C.~Deng, X.~Fang, X.~Han, X.~Wang, L.~Yan, R.~He, Y.~Long, and Y.~Guo, ``{IEEE 802.11be Wi-Fi 7: New Challenges and Opportunities},'' \emph{IEEE Commun. Surveys Tuts.}, vol.~22, no.~4, pp. 2136--2166, 2020.

\bibitem{CFO}
M.~S. Chaudhari, S.~Majhi, and S.~Jain, ``Cnn-attention-dnn design for cfo estimation of non-pilot-assisted ofdm system,'' \emph{IEEE Commun. Lett.}, vol.~27, no.~2, pp. 551--555, 2023.

\end{thebibliography}

\vfill

\end{document}